\documentclass[11pt,a4paper]{article}

\usepackage{amssymb}
\usepackage{graphicx}
\usepackage{bm}

\usepackage{mathrsfs}
\usepackage{cite}
\usepackage{amsmath}

\begin{document}

\title{Construction of mirror pairs  Calabi-Yau orbifolds of the Berglund-Hubsch type}

\author{S.S.Aleshin\,${}^{a}$, A.Belavin\,${}^{b}$,  $\vphantom{\Big(}$
\medskip\\
${}^a${\small{\em Institute for Information Transmission Problems RAS,}}\\
\medskip
{\small{\em 127051, Moscow, Russia,}}\\
${}^b${\small{\em Landau Institute for Theoretical Physics,}}\\
\medskip
{\small{\em 142432 Chernogolovka, Russia}}
\medskip
}

\maketitle

\begin{abstract}
In this paper we have developed general algorithm for finding all orbifolds of Berglund-Hubsch-type Calabi-Yau manifolds and their mirrors. An explicit construction is formulated for finding all admissible deformations and groups defining mirror pairs of orbifolds. Then using our algorithm for one of the Calabi-Yau manifolds, defined by a Fermat-type polynomial, we found all mirror pairs of orbifolds. For this model, for each pair of orbifolds, the number of generations and the number of singlets i.e. particles participating only in gravitational interactions (dark matter particles) were found.



\end{abstract}

\section{Introduction}
\hspace*{\parindent}

A heterotic string in 10-dimensional spacetime was proposed in \cite{Gross:1984dd} as an approach to unifying the Quantum Theory of Gravity and the Grand Unified Theory of Gauge Interactions. After that, in \cite{Candelas:1985en}, the idea was proposed to obtain a heterotic string theory in four-dimensional spacetime by compactifying six dimensions of ten-dimensional spacetime on Calabi–Yau manifolds. Subsequently, D. Gepner put forward the conjecture \cite{Gepner:1987vz, Gepner:1987qi} on the equivalence between the compactification of six dimensions on Calabi–Yau manifolds and compactification on an $N=2$ superconformal theory with central charge $c=9$. This made it possible to construct exactly solvable models of the heterotic string in four-dimensional spacetime.

An important class of exactly solvable heterotic string models, previously considered by D. Gepner \cite{Gepner:1987vz, Gepner:1987qi}, corresponds to products of N = 2 minimal models with total central charge c = 9. The construction of these heterotic string models involves compactification onto Calabi–Yau manifolds, which belong to the special case of CY manifolds of Berglund–Hubsch type.

In work \cite{Belavin:2025yjv}, a method for the explicit construction of heterotic string models compactified on the product of a torus and the Lie algebra $E(8)\times SO(10)$
was developed, as well as on general Calabi–Yau manifolds and their Berglund–Hübsch–type orbifolds. In contrast to the Gepner approach, in work \cite{Belavin:2025yjv}, the compactification is considered for an arbitrary Calabi–Yau manifold of Berglund–Hübsch type. The construction uses the Batyrev–Borisov combinatorial approach \cite{Borisov:2010di}, as well as the construction of vertex operators of physical states using free bosonic and fermionic fields. However, Ref. \cite{Belavin:2025yjv} did not provide a procedure for determining the relevant data for each specific mirror pair of orbifolds, which is required, in particular, to compute the number of dark matter types (i.e., singlets with respect to the gauge group $E(8)\times E(6)$). The present work is devoted to this problem.

In the present work, we have developed a general algorithm for searching for all orbifolds of Berglund–Hübsch–type Calabi–Yau manifolds and their mirror pairs. An explicit construction is formulated for finding all admissible deformations and the groups defining the orbifolds and their mirror pairs. This knowledge is necessary for constructing the heterotic string model for the corresponding Calabi–Yau manifold.

As is known, from a phenomenological point of view, it is important to be able to construct Calabi–Yau manifolds with a topology that determines the number of generations of quarks and leptons to be three. The approach we propose for constructing orbifolds of Calabi–Yau manifolds can help in the search for such models. Using one of the Fermat-type models as an example, we demonstrate our algorithm in action. For this model, for each orbifold, we find the number of generations and the number of particles that participate only in gravitational interactions (singlets), i.e., dark matter particles.

Let us consider the class of Berglund–Hübsch–type Calabi–Yau manifolds, which can be constructed as hypersurfaces in a weighted projective space with weights $ {\bf k} := (k_1,k_2,k_3,k_4,k_5)$
\begin{eqnarray}\label{P}
\mathbb{P}_{{\bf k}}^4 = \Big\{(x_1,...,x_5)\in \mathbb{C}^5\textbackslash \{0\} \,\,\Big |\,\,x_i  \sim \lambda^{k_i} x_i,\,\,\forall\, \lambda\in\mathbb{C}\Big\}
\end{eqnarray}

\noindent
with the following conditions to be satisfied:
\begin{eqnarray}
&&W_0 := \sum_{i=1}^5\prod_{j=1}^5 x_j^{A_{ij}}=0\\
&&\sum_{j=1}^5 A_{ij} k_j = d \label{quasi_homogeneous_polynomial}\\
&&d = \sum_{i=1}^5 k_i\label{c1_0}.
\end{eqnarray}

\noindent
Here $A_{ij}$ is an integer invertible matrix, relation (\ref{quasi_homogeneous_polynomial}) is the quasi-homogeneity condition for the polynomial $W_0$, relation (\ref{c1_0}) is the condition that the hypersurface has a first Chern class equal to zero; this condition is a necessary requirement for Calabi–Yau manifolds.

Let us proceed to the construction of orbifolds for the Calabi–Yau manifold. For this, we define the maximal group of diagonal symmetries
\begin{eqnarray}\label{Aut_0}
&&Aut(A):=\Big\{(\alpha_1,...,\alpha_5)\in(\mathbb{C^*})^5\,\,\Big|\,\, W_0(\alpha_1\cdot x_1,...,\alpha_5\cdot x_5)= W_0(x_1,...,x_5)\Big\}
\end{eqnarray}

\noindent
It can be shown that for any polynomial of one of the atomic types (Fermat, Loop, Chain), as well as for any polynomial that is a combination of such types, it is true that 
$|\alpha_i|=1$.  Then $\alpha_i = e^{2\pi i \theta_i}$, and to satisfy the condition of invariance of the polynomial in (\ref{Aut_0}), it is required that
\begin{eqnarray}
\sum_{j=1}^5 A_{ij}\theta_j \in \mathbb{Z}
\end{eqnarray}

\noindent
from which it follows that the generators of the group 
$Aut(A)$ have the form
\begin{eqnarray}
\Big(e^{2\pi i B_{1j}},..., e^{2\pi i B_{5j}}\Big),\quad j\in\overline{1,5}
\end{eqnarray}

\noindent
where $B_{ij}:= (A^{-1})_{ij}$, then, for an arbitrary $(\alpha_1,...,\alpha_5)\in Aut(A)$ there exist integers $m_j$,  $j\in\overline{1,5}$, such that
\begin{eqnarray}\label{alpha_i}
\alpha_i = exp\Big (2\pi i \sum_{j=1}^5 B_{ij}m_j\Big),\,\quad \forall\,i\in \overline{1,5}
\end{eqnarray}

\noindent
From the definition (\ref{Aut_0}) and the relation (\ref{alpha_i}), it follows that the element $(\alpha_1,...,\alpha_5)\in Aut(A)$ acts as
\begin{eqnarray}
x_i \mapsto \alpha_i\cdot x_i = exp\Big (2\pi i \sum_{j=1}^5 B_{ij}m_j\Big)x_i
\end{eqnarray}

\noindent
Let us consider in $Aut(A)$ the subgroup that preserves $\prod_{i=1}^5x_i$
\begin{eqnarray}\label{SL_pred_final}
G_{\text{adm}}^{\text{max}}=\Big\{p\in Aut(A)\,\,\Big |\,\, p\cdot \prod_{i=1}^5x_i=\prod_{i=1}^5x_i \Big\}
\end{eqnarray}

\noindent
This group is called the maximal admissible group. As its subgroup, the group $G_{\text{adm}}^{\text{max}}$ contains the quantum group
\begin{eqnarray}
J_A := <(e^{\frac{2\pi i}{d}k_1},...,e^{\frac{2\pi i}{d}k_5})>
\end{eqnarray}

\noindent
To define orbifolds, we will consider admissible groups 
$G$ (subgroups of $G_{\text{adm}}^{\text{max}}$) for which\\
$J_A\subseteq G\subseteq G_{\text{adm}}^{\text{max}}$ is satisfied.
Then, for the family $\mathcal{Q}$ of Calabi–Yau manifolds
\begin{eqnarray}
&&\hspace*{-0.20cm}\mathcal{Q}:=\Big\{(x_1,...,x_5)\in \mathbb{P}^4_{\bf k}\,\Big |\, W_0(x)+\sum_{l=1}^{h}\varphi_l
\prod_{i=1}^5 x_i^{S_{li}}=0\Big\}\label{O}
\end{eqnarray}

\noindent
where $\varphi_l$ are moduli of complex structure deformations. Orbifold $X$ is defined as
\begin{eqnarray}
&&X := \mathcal{Q}/\widetilde{G},\quad \text{where}\quad \widetilde{G}:=G/J_A
\end{eqnarray}

\noindent
where $h := h^{2,1}(X)$ is Hodge number, and the admissible deformations of the type $\prod_{i=1}^5 x_i^{S_{li}}$ ($l$ — the number of the deformation) for Berglund–Hübsch–type Calabi–Yau manifolds \cite{Kreuzer:1992bi} are required to satisfy the conditions from the work \cite{Kreuzer:1994np}.

\section{Algorithm for finding Calabi–Yau orbifolds\\ and their mirror pairs}
\hspace*{\parindent}\label{Algorithm}

Let us now define the steps of the algorithm for constructing orbifolds of Calabi–Yau manifolds of Berglund–Hübsch type and their mirror pairs. 

At the first step, we find all admissible deformations of the initial polynomial 
$W_0$. For this, it is necessary to solve the equation
\begin{eqnarray}\label{deformation_condition_1}
\sum_{i=1}^5k_iS_i=d
\end{eqnarray}

\noindent
with one of the restrictions on admissible deformations, depending on the type of the polynomial $W_0$:
\begin{eqnarray}
\text{Fermat}: \quad 0\leq S_i \leq A_{ii}-2,\quad \forall\,i\in\overline{1,5}\label{Ferma_condition}\\
\text{Loop}: \quad 0\leq S_i \leq A_{ii}-1,\quad \forall\,i\in\overline{1,5}
\end{eqnarray} 

\noindent
For the “Chain” type, at least one of the three conditions (\ref{condition 1})–(\ref{condition 3}) must be satisfied:
\begin{eqnarray}\label{condition 1}
&0\le S_{i}\le A_{ii}-\delta_{1i}-1,\quad i\in\overline{1,5}
\end{eqnarray}

\noindent
either the condition
\begin{eqnarray}\label{condition 2}
&&S_{1}=A_{11}-1,\quad S_{2}=0\nonumber\\
&&0\le S_{j}\le A_{jj}-\delta_{3j}-1,\quad j=3,4,5
\end{eqnarray}

\noindent
either the condition
\begin{eqnarray}\label{condition 3}
&&S_{1}=A_{11}-1,\,\, S_{2}=0,\,\, S_{3}= A_{33}-1,\nonumber\\
&& S_{4}=0,\quad 0\le S_{5}\le A_{55}-2,
\end{eqnarray}

\noindent
One may also consider a mixed case.

At the second stage, we determine all elements of the maximal admissible group $G_{\text{adm}}^{\text{max}}$. To this end, we rewrite (\ref{SL_pred_final}) in a form convenient for their determination
\begin{eqnarray}\label{SL_pred_final_Final}
&&\hspace*{-0.3cm} G_{\text{adm}}^{\text{max}}=\Big\{(e^{2\pi ig_1},...,e^{2\pi ig_5})\in(\mathbb{C}^*)^5\,\,\Big |\,\, \sum_{j=1}^5 A_{ij}g_j\in\mathbb{Z},\,\,
\sum_{i=1}^5 g_i \in \mathbb{Z}
\Big\}
\end{eqnarray}

\noindent
An element $(e^{2\pi ig_1},...,e^{2\pi ig_5})$ of the group 
$(\mathbb{C}^*)^5$ belongs to $G_{\text{adm}}^{\text{max}}$
if and only if there exist integers $n_j\in\mathbb{Z}$, such that the following conditions are satisfied
\begin{eqnarray}
&&g_i=\sum_{j=1}^5B_{ij}n_j,\\
&&\sum_{i,j=1}^5 B_{ij}n_j\in\mathbb{Z},\\
&& 0\leq n_j < \max_{i\in\overline{1,5}} A_{ij}\label{group_find}
\end{eqnarray}

At the third step, for each group element
$f_q = (e^{2\pi i g_{q,1}}, \ldots, e^{2\pi i g_{q,5}}) \in G_{\mathrm{adm}}^{\mathrm{max}}$ ($q$ denotes the index of the element in the group) we determine all deformations $\varphi_l \prod_{j=1}^5 x_j^{S_{lj}}$ from the set of all admissible deformations found at the first step that are invariant under the action of $f_q$. Thus, for each group element, one must verify the condition for all admissible deformations
\begin{eqnarray}\label{inv_condition}
\sum_{j=1}^5 g_{q,j}S_{lj}\in \mathbb{Z}
\end{eqnarray}

\noindent
For a group element $f_q$, we denote by $R_q$ the set of all $S_l$ for which the invariance condition (\ref{inv_condition}) is satisfied (hereafter, the elements $S_l$ defining the deformations will also be referred to as deformations). From the resulting list of sets $R_q$, we select all distinct ones and relabel them so that they form a collection $R_i$, $i \in \overline{1, N}$. Next, we determine all possible types of intersections of the sets $R_i$ (let their number be $P$) and add the resulting intersection types to the list $R_i$.

In the fourth step, for each set $R_i$, we use the invariance conditions (\ref{inv_condition}) to identify all elements of the group $G_{\mathrm{adm}}^{\mathrm{max}}$ that act invariantly on the deformations specified by any element of $R_i$. The set of all such group elements corresponding to a given subset of deformations $R_i$ forms a subgroup $G_{\mathrm{adm}}^{i} \subseteq G_{\mathrm{adm}}^{\mathrm{max}}$, which defines the orbifold associated with $R_i$. The resulting pairs $R_i$ and $G_{\mathrm{adm}}^{i}$, for $i \in \overline{1, N+P}$, collectively describe all orbifolds of the model with the polynomial $W_0$.

In the fifth step, for the mirror polynomial defined by the transpose matrix $A^{T}$, we compute all admissible deformations and its maximal admissible diagonal symmetry group $G^{\mathrm{max},T}_{\mathrm{adm}}$, which we call the maximal mirror group.

In the sixth step, for each $R_i$, we determine all mirror deformations $T$ for which the associated pairing takes integral values \cite{Berglund_Hubsch:1998, Krawitz:2009, Belavin:2020xhs}
\begin{eqnarray}\label{pairing}
(S,T):=\sum_{i,j=1}^5 S_{i} B_{ij} T_{j}\in\mathbb{Z}\quad \forall\,S\in R_i.
\end{eqnarray}

\noindent
Denote by $L_i$ the set of all such mirror deformations $T$. For each $L_i$, we find all elements of the group $G^{\text{max},T}_{\text{adm}}$ that act invariantly on every element of $L_i$.These elements form a subgroup
 $G^{T,i}_{\text{adm}} \subseteq G^{\text{max},T}_{\text{adm}}$. The data $L_i$, $G^{T,i}_{\text{adm}}$ define the mirror orbifold to the orbifold $R_i, G^{i}_{\text{adm}}$. 
At this stage, we obtain all mirror pairs $(R_i, L_i)$ of Berglund–Hübsch–type Calabi–Yau orbifolds.

We can now compute the number of generations and the number of matter species (dark matter) that participate only in gravitational interactions, i.e. singlets that do not take part in $E(8)\times E(6)$ gauge interactions. 

In ref. \cite{Belavin:2025yjv}, it was shown that the vertex operators corresponding to particles in the singlet representation of the $E(8)\times E(6)$ algebra can be of three types. The first type consists of pairs of elements of the Batyrev polytope and its mirror whose pairing vanishes.
\begin{eqnarray}\label{singlets}
(S,T)=0,\,\, \forall\, S\in R_{i}\,\, \forall\,T\in L_{i}.
\end{eqnarray}

\noindent
The second type consists of points of the Batyrev polytope, while the third type consists of points of the mirror Batyrev polytope.

The number of generations for the mirror pair with index $i$ is equal to the difference between the numbers of points in the mirror pair of Batyrev polyhedra.
\begin{eqnarray}\label{num_gen}
&&N_{\text{gen}} = |h^{2,1}(Y_i)-h^{2,1}(X_i)|,\\
&& h^{2,1}(X_i)=|R_i|,\quad  h^{2,1}(Y_i) = |L_i|,
\end{eqnarray}

\noindent
where $X_i$ and $Y_i$ are orbifolds associated with the groups $G_{\mathrm{adm}}^i$ and $G_{\mathrm{adm}}^{T,i}$, and with the deformation sets $R_i$ and $L_i$, respectively.

\section{Example}
\hspace*{\parindent}\label{Example}

We apply the algorithm described above as an example to the case of Calabi–Yau manifolds defined by a Fermat-type polynomial with the matrix
$A = \mathrm{diag}(2,4,6,18,36)$ in a weighted projective space with weights $k = (18,9,6,2,1)$.
Let us find the permissible deformations from the conditions:
\begin{eqnarray}
&&\sum_{i=1}^5k_iS_i=d\\
&&0\leq S_i \leq A_{ii}-2,\quad \forall\,i\in\overline{1,5}
\end{eqnarray}

\noindent
We add to them the deformation $x_1 x_2 x_3 x_4 x_5$. In total, the model under consideration has 125 deformations.
\begin{eqnarray}\label{deformation_example}
&&\hspace{-10pt} x_4 x_5^{34},\, \, x_4^{2} x_5^{32},\, \, x_4^{3} x_5^{30},\, \, x_4^{4} x_5^{28},\, \, x_4^{5} x_5^{26},\, \, x_4^{6} x_5^{24},\, \, x_4^{7} x_5^{22},\, \, x_4^{8} x_5^{20},\, \, x_4^{9} x_5^{18},\, \, x_4^{10} x_5^{16},\, \, x_4^{11} x_5^{14},\,\, x_4^{12} x_5^{12},\nonumber\\
&&\hspace{-10pt}  x_4^{13} x_5^{10},\,\, x_4^{14} x_5^{8},\, \, x_4^{15} x_5^{6},\, \, x_4^{16} x_5^{4},\, \, x_3 x_5^{30},\, \, x_3 x_4 x_5^{28},\, \, x_3 x_4^{2} x_5^{26},\, \, x_3 x_4^{4} x_5^{22},\, \, x_3 x_4^{5} x_5^{20},\, \, x_3 x_4^{6} x_5^{18},\nonumber\\
&&\hspace{-10pt}  x_3 x_4^{7} x_5^{16},\, \, x_3 x_4^{8} x_5^{14},\, \, x_3 x_4^{9} x_5^{12},\, \, x_3 x_4^{10} x_5^{10},\, \, x_3 x_4^{11} x_5^{8},\, \, x_3 x_4^{12} x_5^{6},\, \, x_3 x_4^{13} x_5^{4},\, \, x_3 x_4^{14} x_5^{2},\, \, x_3 x_4^{15},\nonumber\\
&&\hspace{-10pt} x_3^{2} x_5^{24},\, \, x_3^{2} x_4 x_5^{22},\, \, x_3^{2} x_4^{2} x_5^{20},\, \, x_3^{2} x_4^{3} x_5^{18},\, \, x_3^{2} x_4^{4} x_5^{16},\, \, x_3^{2} x_4^{5} x_5^{14},\, \, x_3^{2} x_4^{6} x_5^{12},\, \, x_3^{2} x_4^{7} x_5^{10},\, \, x_3^{2} x_4^{8} x_5^{8},\nonumber\\
&&\hspace{-10pt} x_3^{2} x_4^{9} x_5^{6},\,\, x_3^{2} x_4^{10} x_5^{4},\, \, x_3^{2} x_4^{11} x_5^{2},\, \, x_3^{2} x_4^{12},\,\, x_3^{3} x_5^{18},\, \, x_3^{3} x_4 x_5^{16},\, \, x_3^{3} x_4^{2} x_5^{14},\, \, x_3^{3} x_4^{3} x_5^{12},\, \, x_3^{3} x_4^{4} x_5^{10},\, \, x_3^{3} x_4^{5} x_5^{8},\nonumber
\end{eqnarray}

\begin{eqnarray}
&&\hspace{-10pt} x_3^{3} x_4^{6} x_5^{6},\, \, x_3^{3} x_4^{7} x_5^{4},\, \, x_3^{3} x_4^{8} x_5^{2},\, \, x_3^{3} x_4^{9},\, \, x_3 x_4^{3} x_5^{24}, \,\, x_3^{4} x_5^{12},\, \, x_3^{4} x_4 x_5^{10},\, \, x_3^{4} x_4^{2} x_5^{8},\, \, x_3^{4} x_4^{3} x_5^{6},\, \, x_3^{4} x_4^{4} x_5^{4},\, \nonumber\\
&&\hspace{-10pt} x_3^{4} x_4^{5} x_5^{2},\, \, x_3^{4} x_4^{6},\, \, x_2 x_5^{27},\, \, x_2 x_4 x_5^{25},\, \, x_2 x_4^{2} x_5^{23},\, \, x_2 x_4^{3} x_5^{21},\, \, x_2 x_4^{4} x_5^{19},\, \, x_2 x_4^{5} x_5^{17},\, \, x_2 x_4^{6} x_5^{15},\, \, x_2 x_4^{7} x_5^{13},\, \nonumber\\
&&\hspace{-10pt} x_2 x_4^{8} x_5^{11},\, \, x_2 x_4^{9} x_5^{9},\, \, x_2 x_4^{10} x_5^{7},\, \, x_2 x_4^{11} x_5^{5},\, \, x_2 x_4^{12} x_5^{3},\, \, x_2 x_4^{13} x_5,\, \, x_2 x_3 x_5^{21},\, \, x_2 x_3 x_4 x_5^{19},\, \, x_2 x_3 x_4^{2} x_5^{17},\nonumber\\
&&\hspace{-10pt} x_2 x_3 x_4^{3} x_5^{15},\, \, x_2 x_3 x_4^{4} x_5^{13},\, \, x_2 x_3 x_4^{5} x_5^{11},\, \, x_2 x_3 x_4^{6} x_5^{9},\, \, x_2 x_3 x_4^{7} x_5^{7},\, \, x_2 x_3 x_4^{8} x_5^{5},\, \, x_2 x_3 x_4^{9} x_5^{3},\, \, x_2 x_3 x_4^{10} x_5,\nonumber\\
&&\hspace{-10pt} x_2 x_3^{2} x_5^{15},\, \, x_2 x_3^{2} x_4 x_5^{13},\, \, x_2 x_3^{2} x_4^{2} x_5^{11},\,\, x_2 x_3^{2} x_4^{3} x_5^{9},\, \, x_2 x_3^{2} x_4^{4} x_5^{7},\, \, x_2 x_3^{2} x_4^{5} x_5^{5},\, \, x_2 x_3^{2} x_4^{6} x_5^{3},\, \, x_2 x_3^{2} x_4^{7} x_5,\, \nonumber\\
&&\hspace{-10pt} x_2 x_3^{3} x_5^{9},\, \, x_2 x_3^{3} x_4 x_5^{7},\, \, x_2 x_3^{3} x_4^{2} x_5^{5},\, \, x_2 x_3^{3} x_4^{3} x_5^{3},\, \, x_2 x_3^{3} x_4^{4} x_5,\, \, x_2 x_3^{4} x_5^{3},\, \, x_2 x_3^{4} x_4 x_5,\, \, x_2^{2} x_5^{18},\, \, x_2^{2} x_4 x_5^{16},\nonumber\\
&&\hspace{-10pt} x_2^{2} x_4^{2} x_5^{14},\, \, x_2^{2} x_4^{3} x_5^{12},\, \, x_2^{2} x_4^{4} x_5^{10},\, \, x_2^{2} x_4^{5} x_5^{8},\, \, x_2^{2} x_4^{6} x_5^{6},\, \, x_2^{2} x_4^{7} x_5^{4},\, \, x_2^{2} x_4^{8} x_5^{2},\, \, x_2^{2} x_4^{9},\, \, x_2^{2} x_3 x_5^{12},\, \, x_2^{2} x_3 x_4 x_5^{10},\nonumber\\
&&\hspace{-10pt} x_2^{2} x_3 x_4^{2} x_5^{8},\, \, x_2^{2} x_3 x_4^{3} x_5^{6},\, \, x_2^{2} x_3 x_4^{4} x_5^{4},\, \, x_2^{2} x_3 x_4^{5} x_5^{2},\, \, x_2^{2} x_3 x_4^{6},\, \, x_2^{2} x_3^{2} x_5^{6},\, \, x_2^{2} x_3^{2} x_4 x_5^{4},\, \, x_2^{2} x_3^{2} x_4^{2} x_5^{2},\, \, x_2^{2} x_3^{2} x_4^{3},\nonumber\\
&&\hspace{-10pt} x_2^{2} x_3^{3},\, \,x_1 x_2 x_3 x_4 x_5\nonumber
\end{eqnarray}

From conditions (\ref{SL_pred_final_Final})–(\ref{group_find}), all 864 elements of the maximal admissible group $G_{\text{adm}}^{\text{max}}$ were found. Then, for each group element $(e^{2\pi i g_{q,1}}, \ldots, e^{2\pi i g_{q,5}}) \in G_{\text{adm}}^{\text{max}}$, the set $R_q$ of all deformations satisfying the invariance condition (\ref{inv_condition}) was determined.

It turned out that for the model under consideration there are only $N = 16$ distinct types of sets, which we denote by $R_i$. Table 1 lists the group elements $g_i \in G_{\text{adm}}^{\text{max}}$ that define the deformation sets $R_i$. Each element $g_i$ has, as the set of all deformations on which it acts invariantly, the deformation set $R_i$.

\begin{table}[h]
\centering
    \begin{minipage}{0.45\textwidth}
        \centering
        \label{tab:first}
\begin{tabular}{rcl}
\hline
\multicolumn{1}{|l|}{$\,{\bf i}$} & \multicolumn{1}{l|}{$\qquad\qquad\quad {\bf g_i \in\,G_{\text{adm}}^{\text{max}}}$}\\ \hline
\multicolumn{1}{|l|}{1} & \multicolumn{1}{l|}{$(e^{2\pi i \frac{1}{2}},e^{2\pi i \frac{1}{4}},e^{2\pi i \frac{1}{6}},e^{2\pi i \frac{1}{18}},e^{2\pi i \frac{1}{36}})$}              \\ \hline
\multicolumn{1}{|l|}{2} & \multicolumn{1}{l|}{$(1,1,1,e^{2\pi i \frac{1}{18}},e^{2\pi i \frac{17}{18}})$}              \\ \hline
\multicolumn{1}{|l|}{3} & \multicolumn{1}{l|}{$(1,1,1,e^{2\pi i \frac{1}{9}},e^{2\pi i \frac{8}{9}})$}              \\ \hline
\multicolumn{1}{|l|}{4} & \multicolumn{1}{l|}{$(1,1,1,e^{2\pi i \frac{1}{6}},e^{2\pi i \frac{5}{6}})$}              \\ \hline
\multicolumn{1}{|l|}{5} & \multicolumn{1}{l|}{$(1,1,e^{2\pi i \frac{1}{6}},1,e^{2\pi i \frac{5}{6}})$}              \\ \hline
\multicolumn{1}{|l|}{6} & \multicolumn{1}{l|}{$(1,1,e^{2\pi i \frac{1}{6}},e^{2\pi i \frac{1}{18}},e^{2\pi i \frac{7}{9}})$}              \\ \hline
\multicolumn{1}{|l|}{7} & \multicolumn{1}{l|}{$(1,1,e^{2\pi i \frac{1}{6}},e^{2\pi i \frac{1}{6}},e^{2\pi i \frac{2}{3}})$}              \\ \hline
\multicolumn{1}{|l|}{8} & \multicolumn{1}{l|}{$(1,1,e^{2\pi i \frac{1}{6}},e^{2\pi i \frac{2}{9}},e^{2\pi i \frac{11}{18}})$}              \\ \hline
                        &                                                         
\end{tabular}
    \end{minipage}%
    \begin{minipage}{0.45\textwidth}
        \centering
        \label{tab:second}
\begin{tabular}{rcl}
\hline
\multicolumn{1}{|l|}{$\,{\bf i}$} & \multicolumn{1}{l|}{$\qquad\qquad {\bf g_i \in\,G_{\text{adm}}^{\text{max}}}$}\\ \hline
\multicolumn{1}{|l|}{9} & \multicolumn{1}{l|}{$(1,e^{2\pi i \frac{1}{4}},1,1,e^{2\pi i \frac{3}{4}})$}              \\ \hline
\multicolumn{1}{|l|}{10} & \multicolumn{1}{l|}{$(1,e^{2\pi i \frac{1}{4}},1,e^{2\pi i \frac{1}{18}},e^{2\pi i \frac{25}{36}})$}              \\ \hline
\multicolumn{1}{|l|}{11} & \multicolumn{1}{l|}{$(1,e^{2\pi i \frac{1}{4}},1,e^{2\pi i \frac{1}{9}},e^{2\pi i \frac{23}{36}})$}              \\ \hline
\multicolumn{1}{|l|}{12} & \multicolumn{1}{l|}{$(1,e^{2\pi i \frac{1}{4}},1,e^{2\pi i \frac{1}{6}},e^{2\pi i \frac{7}{12}})$}              \\ \hline
\multicolumn{1}{|l|}{13} & \multicolumn{1}{l|}{$(1,e^{2\pi i \frac{1}{4}},e^{2\pi i \frac{1}{6}},1,e^{2\pi i \frac{7}{12}})$}              \\ \hline
\multicolumn{1}{|l|}{14} & \multicolumn{1}{l|}{$(1,e^{2\pi i \frac{1}{4}},e^{2\pi i \frac{1}{6}},e^{2\pi i \frac{1}{18}},e^{2\pi i \frac{19}{36}})$}              \\ \hline
\multicolumn{1}{|l|}{15} & \multicolumn{1}{l|}{$(1,e^{2\pi i \frac{1}{4}},e^{2\pi i \frac{1}{6}},e^{2\pi i \frac{1}{6}},e^{2\pi i \frac{5}{12}})$}              \\ \hline
\multicolumn{1}{|l|}{16} & \multicolumn{1}{l|}{$(1,e^{2\pi i \frac{1}{4}},e^{2\pi i \frac{1}{6}},e^{2\pi i \frac{2}{9}},e^{2\pi i \frac{13}{36}})$}              \\ \hline
                        &                                               
\end{tabular}
    \end{minipage}%
    
\caption{Elements of the group $g_i \in G_{\text{adm}}^{\text{max}}$ defining the deformation sets $R_i$.}
\end{table}


For example, the orbifold $X_{15}$ is defined by the set of deformations $R_{15}$, consisting of $h^{2,1}(X_{15}) = 29$ elements:

\begin{eqnarray}
&&(0, 0, 0, 3, 30),\,\,\,\,\,(0, 0, 0, 6, 24),\,\,\,\,\,(0, 0, 0, 9, 18),\,\,\,\,\,(0, 0, 0, 12, 12),\,\,\,\,\,(0, 0, 0, 15, 6),\nonumber\\
&&(0, 0, 1, 1, 28),\,\,\,\,\,(0, 0, 1, 4, 22),\,\,\,\,\,(0, 0, 1, 7, 16),\,\,\,\,\,(0, 0, 1, 10, 10),\,\,\,\,\,(0, 0, 2, 2, 20),\nonumber\\
&&(0, 0, 2, 5, 14),\,\,\,\,\,(0, 0, 2, 8, 8),\,\,\,\,\,\,\,\,(0, 0, 2, 11, 2),\,\,\,\,\,(0, 0, 3, 0, 18),\,\,\,\,\,\,\,\,(0, 0, 3, 3, 12),\nonumber\\
&&(0, 0, 3, 6, 6),\,\,\,\,\,\,\,\,(0, 0, 3, 9, 0),\,\,\,\,\,\,\,\,(0, 0, 4, 1, 10),\,\,\,\,\,(0, 0, 4, 4, 4),\,\,\,\,\,\,\,\,\,\,\,(0, 2, 0, 0, 18),\nonumber\\
&&(0, 2, 0, 3, 12),\,\,\,\,\,(0, 2, 0, 6, 6),\,\,\,\,\,\,\,\,(0, 2, 0, 9, 0),\,\,\,\,\,\,\,\,(0, 2, 1, 1, 10),\,\,\,\,\,\,\,(0, 2, 1, 4, 4),\nonumber\\
&&   (0, 2, 2, 2, 2),\,\,\,\,\,\,\,\,(0, 2, 3, 0, 0),\,\,\,\,\,\,\,\,(0, 0, 1, 13, 4),\,\,\,\,\,(1, 1, 1, 1, 1).\nonumber
\end{eqnarray}

\noindent
Then we find all possible types of intersections of the sets $R_i$. They are listed in column $R_{16+i}$ of Table 2.

In the next step, for each set of deformations $R_i$, we use condition (\ref{inv_condition}) to determine all elements of the maximal admissible group that act invariantly on any deformation specified by an element of $R_i$. The set of all such group elements for each subset of deformations $R_i$ forms a subgroup $G^i_{\text{adm}} \subseteq G_{\text{adm}}^{\text{max}}$, which defines the orbifold associated with $R_i$. The number of deformations in $R_i$ $\bigl(h^{2,1}(X_i)\bigr)$ and
the orders of the groups $G^i_{\text{adm}}$ are given in Table 2.

\begin{table}[h]
\centering
\begin{tabular}{|l|l|l|l|l|l|}
\hline
$\,\,{\bf i}$ & ${\bf h^{2,1}(X_i)}$ & ${\bf |G_{\text{adm}}^i|}$ & $\qquad {\bf R_{16+i}}$ & ${\bf h^{2,1}(Y_i)}$ & ${\bf |G_{\text{adm}}^{T,i}|}$ \\ \hline
1   & \,\,\,\,\,\,125     & \,\,\,\,\,\,36      & $R_2\cap R_5\cap R_9$      &        \,\,\,\,\,\,\,\,9 &   \,\,\,\,864        \\ \hline
2   & \,\,\,\,\,\,\,\,22      & \,\,\,216     &\,\,\,\,\,\,\,$R_8\cap R_{12}$       &        \,\,\,\,\,\,51 & \,\,\,\,144          \\ \hline
3   & \,\,\,\,\,\,\,\,43      & \,\,\,108     &$R_4\cap R_6\cap R_9$       &        \,\,\,\,\,\,27 & \,\,\,\,288          \\ \hline
4   & \,\,\,\,\,\,\,\,64      & \,\,\,\,\,\,72     &\,\,\,\,\,\,\,$R_5\cap R_{10}$       &        \,\,\,\,\,\,17 &  \,\,\,\,432         \\ \hline
5   & \,\,\,\,\,\,\,\,23      & \,\,\,216     &\,\,\,\,\,\,\,$R_4\cap R_{14}$       &\,\,\,\,\,\,45         &  \,\,\,\,144         \\ \hline
6   & \,\,\,\,\,\,\,\,62      & \,\,\,\,\,\,72     &\,\,\,\,\,\,\,$R_6\cap R_{11}$       &\,\,\,\,\,\,15         &  \,\,\,\,432         \\ \hline
7   & \,\,\,\,\,\,\,\,22      & \,\,\,216     &\,\,\,\,\, $R_6\cap R_{9}$      &        \,\,\,\,\,\,43 &  \,\,\,\,144         \\ \hline
8   & \,\,\,\,\,\,\,\,67      & \,\,\,\,\,\,72     &\,\,\,\,\,\,\,$R_2\cap R_{13}$       &        \,\,\,\,\,\,15 & \,\,\,\,432          \\ \hline
9   &\,\,\,\,\,\,\,\,64      &\,\,\,\,\,\,72       & \,\,\,\,\,\,\,$R_6\cap R_{10}$      &        \,\,\,\,\,\,13 & \,\,\,\,432          \\ \hline
10   &\,\,\,\,\,\,\,\,25       & \,\,\,216     &\,\,\,\,\,\,\,$R_4\cap R_{9}$       &        \,\,\,\,\,\,39 & \,\,\,\,144          \\ \hline
11   & \,\,\,\,\,\,\,\,22      & \,\,\,216     & \,\,\,\,\,\,\,$R_6\cap R_{12}$       &  \,\,\,\,\,\,39       & \,\,\,\,144 \\ \hline
   12& \,\,\,\,\,\,\,\,75      & \,\,\,\,\,\,72     &\,\,\,\,\,\,\,$R_2\cap R_{9}$      &        \,\,\,\,\,\,13 &  \,\,\,\,432 \\ \hline
13   & \,\,\,\,\,\,\,\,22      &  \,\,\,216    &\,\,\,\,\,\,\,$R_8\cap R_{9}$ &        \,\,\,\,\,\,36 &  \,\,\,\,144 \\ \hline
14   & \,\,\,\,\,\,\,\,85      & \,\,\,\,\,\,72     &\,\,\,\,\,\,\,$R_2\cap R_{5}$ &        \,\,\,\,\,\,12 &  \,\,\,\,432 \\ \hline
15   & \,\,\,\,\,\,\,\,29      &  \,\,\,216    &\,\,\,\,\,\,\,$R_4\cap R_{6}$      &        \,\,\,\,\,\,34 & \,\,\,\,144  \\ \hline
16   &  \,\,\,\,\,\,\,\,66     & \,\,\,\,\,\,72     &\,\,\,\,\,\,\,$R_5\cap R_{9}$       &        \,\,\,\,\,\,12 & \,\,\,\,432 \\ \hline
\end{tabular}
\caption{The numbers of deformations in the sets $R_i$ and $L_i$ ($h^{2,1}(X_i)$ and $h^{2,1}(Y_i)$) and the orders of the groups $G^i_{\text{adm}}$ and $G^{T,i}_{\text{adm}}$.}
\end{table}

\noindent
Note that since in Fermat-type models $A = A^{T}$, the set of all mirror deformations coincides with the set of admissible original deformations. Likewise, the set of orbifolds for the mirror model coincides with the set of orbifolds for the original model.

Using condition (\ref{pairing}), we determine the subsets of mirror deformations $L_i$. Then, proceeding analogously to the steps used to compute the groups $G^{i}_{\text{adm}}$, we find all groups $G^{T,i}_{\text{adm}}$, $i \in \overline{1,16}$. The number of deformations in $L_i$ $\bigl(h^{2,1}(Y_i)\bigr)$ and
the orders of the groups $G^{T,i}_{\text{adm}}$ are given in Table 2.


For example, the orbifold $Y_{15}$ is defined by the set of deformations $L_{15}$, consisting of $h^{2,1}(Y_{15}) = 34$ elements:

\begin{eqnarray}
&&(1, 1, 1, 1, 1), \,\,\,\,\,\,\,\,\,(0, 0, 0, 2, 32)\,\,\,\,\,\,\,\,\,\,\,(0, 0, 0, 4, 28),\,\,\,\,(0, 0, 0, 6, 24),\,\,\,\,(0, 0, 0, 8, 20),\nonumber\\
&&(0, 0, 0, 10, 16),\,\,\,(0, 0, 0, 12, 12),\,\,\,\,\,(0, 0, 0, 14, 8),\,\,\,\,(0, 0, 0, 16, 4),\,\,\,\,(0, 0, 2, 0, 24)\nonumber\\
&&(0, 0, 2, 2, 20),\,\,\,\,\,\,(0, 0, 2, 4, 16),\,\,\,\,\,\,\,\,(0, 0, 2, 6, 12),\,\,\,\,(0, 0, 2, 8, 8),\,\,\,\,\,\,\,(0, 0, 2, 10, 4),\nonumber\\
&&(0, 0, 2, 12, 0),\,\,\,\,\,\,(0, 0, 4, 0, 12),\,\,\,\,\,\,\,\,(0, 0, 4, 2, 8),\,\,\,\,\,\,\,(0, 0, 4, 4, 4),\,\,\,\,\,\,\,(0, 0, 4, 6, 0),\nonumber\\
&&(0, 1, 1, 1, 19),\,\,\,\,\,\,(0, 1, 1, 3, 15),\,\,\,\,\,\,\,\,(0, 1, 1, 5, 11),\,\,\,\,(0, 1, 1, 7, 7),\,\,\,\,\,\,\,(0, 1, 1, 9, 3)\nonumber\\
&&(0, 1, 3, 1, 7),\,\,\,\,\,\,\,\,\,(0, 1, 3, 3, 3),\,\,\,\,\,\,\,\,\,\,\,(0, 2, 0, 0, 18),\,\,\,\,(0, 2, 0, 2, 14),\,\,\,\,(0, 2, 0, 4, 10),\nonumber\\
&&(0, 2, 0, 6, 6),\,\,\,\,\,\,\,\,\,(0, 2, 0, 8, 2),\,\,\,\,\,\,\,\,\,\,\,(0, 2, 2, 0, 6),\,\,\,\,\,\,\,(0, 2, 2, 2, 2)\nonumber
\end{eqnarray}

\noindent
and the corresponding group defining the orbifold has order
$|G^{T,15}_{\text{adm}}|=144$

Comparing the sets of mirror deformations $L_i$ and the sets of intersection types, we see that for these sets the relation $L_i = R_{16+i}$ holds. Consequently, the group defining the orbifold $R_{16+i}$ coincides with the group for the set of mirror deformations $L_i$ found earlier. It follows that the complete set of mirror pairs for the model under consideration consists of 
\begin{eqnarray}
(R_i,L_i)\quad\text{and}\quad (L_i,R_i),\quad i\in\overline{1,16}
\end{eqnarray}

\noindent
The orders of the groups for all orbifolds are listed in Table 2.

Let us now determine the number of generations and the number of singlets for theories compactified on the Calabi–Yau orbifolds we have found. For singlets of the first type, one needs to find the number of deformations and mirror deformations that satisfy the singlet condition, i.e. whose pairing is equal to zero. The numbers of singlets of the second and third types can be found from (\ref{singlets1}) and (\ref{singlets2}), respectively, as the numbers of lattice points in the corresponding Batyrev polytopes. For Calabi–Yau orbifolds of Fermat type, they are obtained from the relations
\begin{eqnarray}
&&N_{\text{singlets}}=|R_i|,\label{singlets1}\\
&&N_{\text{singlets}}=|L_i|.\label{singlets2}
\end{eqnarray}

\noindent
The number of generations is obtained from (\ref{num_gen}). The results for the numbers of generations $N_{\text{gen}}$ and singlets $N_{\text{singlets}}$ for the orbifolds of our example are collected in Table 3.

\begin{table}[h]
\centering
    \begin{minipage}{.36\textwidth}
        \centering
      
        \label{tab:first}
\begin{tabular}{rcl}
\hline
\multicolumn{1}{|l|}{${\bf i}$} & \multicolumn{1}{l|}{${\bf N_\text{gen}}$} & \multicolumn{1}{l|}{\,\,\,\,${\bf N_{\text{singlets}}}$} \\ \hline
\multicolumn{1}{|l|}{1}& \multicolumn{1}{l|}{116}                      & \multicolumn{1}{l|}{(6,\,125,\,9)}  \\ \hline
\multicolumn{1}{|l|}{2} & \multicolumn{1}{l|}{29}                      & \multicolumn{1}{l|}{(19,\,22,\,51)}                \\ \hline
\multicolumn{1}{|l|}{3} & \multicolumn{1}{l|}{16}                      & \multicolumn{1}{l|}{(15,\,43,\,27)}                \\ \hline
\multicolumn{1}{|l|}{4} & \multicolumn{1}{l|}{47}                      & \multicolumn{1}{l|}{(11,\,64,\,17)}                \\ \hline
\multicolumn{1}{|l|}{5} & \multicolumn{1}{l|}{22}                      & \multicolumn{1}{l|}{(11,\,23,\,45)}                \\ \hline
\multicolumn{1}{|l|}{6} & \multicolumn{1}{l|}{47}                      & \multicolumn{1}{l|}{(6,\,62,\,15)}                \\ \hline
\multicolumn{1}{|l|}{7} & \multicolumn{1}{l|}{21}                      & \multicolumn{1}{l|}{(9,\,22,\,43)}                \\ \hline
\multicolumn{1}{|l|}{8} & \multicolumn{1}{l|}{52}                      & \multicolumn{1}{l|}{(19,\,67,\,15)}                \\ \hline
                        &                                                         &  
\end{tabular}
    \end{minipage}%
    \begin{minipage}{.36\textwidth}
        \centering
        \label{tab:second}
\begin{tabular}{rcl}
\hline
\multicolumn{1}{|l|}{\,${\bf i}$} & \multicolumn{1}{l|}{${\bf N_{\text{gen}}}$} & \multicolumn{1}{l|}{\,\,\,\,${\bf N_{\text{singlets}}}$} \\ \hline
\multicolumn{1}{|l|}{9}& \multicolumn{1}{l|}{51}                      & \multicolumn{1}{l|}{(8,\,64,\,13)}  \\ \hline
\multicolumn{1}{|l|}{10} & \multicolumn{1}{l|}{14}                      & \multicolumn{1}{l|}{(6,\,25,\,39)}                \\ \hline
\multicolumn{1}{|l|}{11} & \multicolumn{1}{l|}{17}                      & \multicolumn{1}{l|}{(4,\,22,\,39)}                \\ \hline
\multicolumn{1}{|l|}{12} & \multicolumn{1}{l|}{62}                      & \multicolumn{1}{l|}{(2,\,75,\,13)}                \\ \hline
\multicolumn{1}{|l|}{13} & \multicolumn{1}{l|}{14}                      & \multicolumn{1}{l|}{(12,\,22,\,36)}                \\ \hline
\multicolumn{1}{|l|}{14} & \multicolumn{1}{l|}{73}                      & \multicolumn{1}{l|}{(6,\,85,\,12)}                \\ \hline
\multicolumn{1}{|l|}{15} & \multicolumn{1}{l|}{5}                      & \multicolumn{1}{l|}{(13,\,29,\,34)}                \\ \hline
\multicolumn{1}{|l|}{16} & \multicolumn{1}{l|}{54}                      & \multicolumn{1}{l|}{(4,\,66,\,12)}                \\ \hline
                        &                                                         &  
\end{tabular}
    \end{minipage}
\caption{The number of generations $N_{\text{gen}}$ and singlets $N_{\text{singlets}}$. The position index of a number in the column vector $N_{\text{singlets}}$ corresponds to the type of singlet.
}
\end{table}

\section{Conclusion}
\hspace*{\parindent}

In this paper, we developed an algorithm for obtaining admissible deformations and the groups that define all Berglund–Hübsch type Calabi–Yau orbifolds and their mirror pairs. The algorithm we propose can be used to search for Calabi–Yau manifolds with a topology that gives the number of quark and lepton generations equal to three, which is important from a phenomenological point of view. Using one of the Fermat models as an example, we demonstrated the algorithm in action: all orbifolds and their mirror pairs for the model under consideration were found. In addition, for each orbifold we computed the number of generations of leptons and quarks, as well as the number of singlet particles, i.e., particles that do not participate in the gauge interaction. Since such particles interact only gravitationally, they can be identified as dark matter particles. The results obtained in this article have been published in the journal \cite{Aleshin_Belavin: 2026}.

\hspace*{1pt}

\section*{Acknowledgments}
\hspace*{\parindent}

A. Belavin acknowledges D. Gepner M. Jibladze G. Koshevoy and A. Litvinov for the helpful discussions. The research of S. Aleshin was carried out within the state assignment of Ministry of Science and Higher Education of the Russian Federation for IITP RAS. The work of A. Belavin has been supported by the state assignment FFWR-2024-0012.

\end{document}